\documentclass[a4paper,12pt]{article}
\usepackage[T1]{fontenc}
\usepackage[backend=biber,style=phys,biblabel=brackets,chaptertitle=false,pageranges=false,eprint=true]{biblatex}
\usepackage[affil-it]{authblk}
\usepackage{setspace}
\usepackage{amsmath}
\usepackage{amssymb}
\usepackage{bm}
\usepackage{booktabs}
\usepackage[nopatch=item]{microtype}
\usepackage[colorlinks, allcolors=blue]{hyperref}
\usepackage{graphicx}
\usepackage{tikzexternal}
\title{SPARSE: Scattering Poles and Amplitudes from Radial Schr\"odinger Equations}

\author{%
	Roberto Bruschini%
	\footnote{%
		\href{mailto:roberto.bruschini@tum.de}{roberto.bruschini@tum.de}%
	}%
}
\affil{Department of Physics, Technical University of Munich\\   
Garching, 85748, Germany}
\affil{Department of Physics, The Ohio State University\\
Columbus, Ohio 43210, USA}
\date{}

\begin{document}

\maketitle

\begin{center}
\textit{GitHub repository link:} \href{https://github.com/Robrusch/SPARSE}{github.com/Robrusch/SPARSE}
\end{center}

\begin{abstract}
We introduce an algorithm for the solution of a system of radial Schr\"odinger equations describing the inelastic scattering of particles with spin in a partial wave with definite total angular momentum.
The system of differential equations is approximated as an ordinary linear nonhomogeneous system using the finite difference method.
Dirichlet boundary conditions are imposed at the origin and at an arbitrary large radius.
The numerical wavefunction is calculated using LAPACK general banded LU decomposition routines.
The $K$-matrix for physical energies is calculated from the numerical solutions of the system by least-square fit to
the analytical real solutions at large distances.
The scattering amplitudes for physical energies is then calculated from the $K$-matrix.
Scattering poles in the lower half of the unphysical Riemann sheet are calculated by numerical extrapolation of the scattering amplitudes using the AAA algorithm for rational approximation.
\end{abstract}

\section{Introduction}
\label{sec:intro}

The Schr\"odinger equation applies to scattering states just as well as to bound states.
In fact, the Schr\"odinger equation provides the foundation of many approximate analytical methods to calculate scattering poles and amplitudes using nonrelativistic quantum mechanics.
But the advantages of calculating scattering poles and amplitudes directly from the Schr\"odinger equation are sometimes overlooked.

The most obvious benefit of calculating scattering states directly from the Schr\"odinger equation is that it does not require any secondary approximation that may limit the validity of the results.
The scattering wavefunctions are completely determined by the Schr\"odinger equation, which in turn is completely determined by the potential and the masses of the degrees of freedom.
Therefore, the scattering poles and amplitudes obtained in this way are just as good as the model or framework in which the Schr\"odinger equation is defined.
That is most advantageous for those cases where resonant line shapes overlap with one another or with the energy thresholds where scattering channels open up.
There is no need for a special treatment of esoteric structures like dips and cusps in the scattering amplitudes, since they are automatically described by the Schr\"odinger equation.

A system of Schr\"odinger equations can be used to describe the quantum mechanical scattering of particles with spin in systems where different final states are possible after a collision at nonrelativistic center-of-mass energies.
One notable example is nucleon-nucleon scattering, where inelastic effects start to play a role at relatively low scattering momenta ($NN\to NN, N \Delta, N N^\prime,\Delta\Delta \dots$).

In this paper, we introduce a simple yet effective algorithm for calculating scattering poles and amplitudes directly from a system of radial Schr\"odinger equations.
Our method is to solve the system using a first-order finite-difference method (see Chapter~3 of Reference~\cite{Kal02}).
The crudeness of the finite-difference method is offset by its efficiency, which makes it the gold standard for numerically solving a differential equation with minimal effort.
More sophisticated methods include propagation algorithms \cite{Man86,But96} and basis expansions \cite{Ors80,Lil82} (see Reference~\cite{Kal02} for a review).

In phenomenological applications, there are many situations in which one has to solve a system of Schr\"odinger equations several times to obtain a definite prediction.
For instance, one may need to fit the parameters of some potential model, or one may want to estimate the error on the calculated spectrum from the uncertainty of the potential parameters using a bootstrap method.
In those cases, it is desirable to have an algorithm that solves the system of Scrh\"odinger equations as efficiently as possible.
That is the main purpose of the SPARSE algorithm.
The extreme efficiency of the finite-difference method allows SPARSE to solve systems with tens of coupled Schr\"odinger equations using only modest computing power and time.

The rest of the paper is structured as follows.
In Section~\ref{sec:equation}, we derive the system of radial Schr\"odinger equations describing an inelastic two-body scattering process of particles with spin in a partial wave with definite total-angular-momentum quantum numbers.
In Section~\ref{sec:wavefunc}, we reduce the system of radial Schr\"odinger equations down to a nonhomogeneous linear system.
In Section~\ref{sec:kmatrix}, we calculate the reactance matrix $K$ for the scattering process by fitting the numerical wavefunctions to the analytic ones for scattering states with real boundary conditions.
In Section~\ref{sec:scattering}, we review the calculation of scattering poles and amplitudes.
In Section~\ref{sec:showcase}, we showcase the algorithm using a simple example that serves also as a tutorial for SPARSE.
Finally, we summarize these results in Section~\ref{sec:summary}.

\section{System of radial Schr\"odinger equations}
\label{sec:equation}

A two-body elastic or inelastic scattering process at nonrelativistic energies can be described by a system of coupled Schr\"odinger equations.
The system can be written explicitly as
\begin{equation}
\sum_{b=1}^\mathcal{N}\biggl[\delta_{ab}\biggl(\frac{p_1^2}{2 m_{1,a}} + \frac{p_2^2}{2 m_{2,a}}\biggr) + V_{ab}(\bm{r}_1 - \bm{r}_2) - \delta_{ab}E\biggr] \psi_b(\bm{r}_1,\bm{r}_2)=0,
\label{eq:schr2bodexplicit}
\end{equation}
where $a$ and $b$ are integers labeling the different Schr\"odinger equations in the system, also referred to as \emph{channels}, $\mathcal{N}$ is the total number of channels, $\bm{r}_1$ and $\bm{r}_2$ are the particles' coordinates, $\bm{p}_1=-i\bm{\nabla}_{\bm{r}_1}$ and $\bm{p}_2=-i\bm{\nabla}_{\bm{r}_2}$ are their conjugate momenta, $m_{1,a}$ and $m_{2,a}$ are the masses of the interacting particles in the channel $a$, $V_{ab}(\bm{r}_1 - \bm{r}_2)$ is the potential coupling the channel $b$ to the channel $a$, $\psi_a(\bm{r}_1, \bm{r}_2)$ is the two-body wavefunction in the channel $a$, and $E$ is the total energy.
Note that we use natural units, $\hbar= c = 1$, and that unitarity requires $V_{ab}^\dag(\bm{r}_1 - \bm{r}_2) = V_{ba}(\bm{r}_1 - \bm{r}_2)$.
In general, Equation~\eqref{eq:schr2bodexplicit} describes interacting particles with spin.
Each wavefunction $\psi_a(\bm{r}_1, \bm{r}_2)$ contains the spin states of the interacting particles in the channel $a$ and each potential $V_{ab}(\bm{r}_1 - \bm{r}_2)$ is an operator mapping spin vectors in the channel $b$ to spin vectors in the channel $a$.
Note also that the system in Equation~\eqref{eq:schr2bodexplicit} describes a generally inelastic scattering process where the state of the interacting particles can change during the collision. Therefore, in general one has $m_{1,a}\neq m_{1,b}$ and $m_{2,a}\neq m_{2,b}$ for $a\neq b$.
Using matrix notation, the system in Equation~\eqref{eq:schr2bodexplicit} can be written more concisely as
\begin{equation}
\biggl[\frac{p_1^2}{2 m_1} + \frac{p_2^2}{2 m_2} + V(\bm{r}_1 - \bm{r}_2) - E\biggr] \psi(\bm{r}_1,\bm{r}_2)=0,
\label{eq:schr2bod}
\end{equation}
where $V(\bm{r}_1 - \bm{r}_2)$ is a $\mathcal{N}\times\mathcal{N}$ matrix of potentials and $\psi(\bm{r}_1, \bm{r}_2)$ is a vector wavefunction.
The $m_1$ and $m_2$ are diagonal mass matrices whose elements are the masses of the interacting particles in the various channels.
Each channel, taken in isolation, describes an elastic scattering process.
The couplings between channels with different masses, which are mediated by the off-diagonal elements of $V(\bm{r}_1 - \bm{r}_2)$, produce inelastic effects.

We wish to reduce Equation~\eqref{eq:schr2bod} down to a system of differential equations in a single vector variable. Let us introduce the relative position
\begin{equation}
\bm{r} = \bm{r}_1 - \bm{r}_2
\label{eq:relr}
\end{equation}
and the center-of-average-mass position
\begin{equation}
\bm{R}= \frac{\overline{m}_1 \bm{r}_1 + \overline{m}_2 \bm{r}_2}{\overline{m}_1 + \overline{m}_2},
\end{equation}
where $\overline{m}_1 = \sum_{a=1}^\mathcal{N} m_{1,a} / \mathcal{N}$ and $\overline{m}_2 = \sum_{a=1}^\mathcal{N} m_{2,a} / \mathcal{N}$ are the average of the masses in the diagonal matrices $m_1$ and $m_2$.
We also introduce the relative momentum
\begin{equation}
\bm{p}= \frac{\overline{m}_2 \bm{p}_1 - \overline{m}_1 \bm{p}_2}{\overline{m}_1 + \overline{m}_2}
\end{equation}
and the total momentum
\begin{equation}
\bm{P} = \bm{p}_1 + \bm{p}_2,
\label{eq:totp}
\end{equation}
which are the momenta conjugate to $\bm{r}$ and $\bm{R}$, that is, $\bm{p}=-i\bm{\nabla}_{\bm{r}}$ and $\bm{P}=-i\bm{\nabla}_{\bm{R}}$.
Expressing $\bm{r}_1$, $\bm{r}_2$, $\bm{p}_1$, and $\bm{p}_2$ inside Equation \eqref{eq:schr2bod} in terms of $\bm{r}$, $\bm{R}$, $\bm{p}$, and $\bm{P}$ using Equations~\eqref{eq:relr}-\eqref{eq:totp}, we obtain
\begin{multline}
\biggl[\frac{p^2}{2 \mu} + \biggl(\frac{\overline{m}_1}{m_1} -\frac{\overline{m}_2}{m_2}\biggr) \frac{\bm{p}\cdot\bm{P}}{\overline{m}_1 + \overline{m}_2} + \biggl(\frac{\overline{m}_1^2}{m_1} +\frac{\overline{m}_2^2}{m_2}\biggr) \frac{P^2}{2(\overline{m}_1 + \overline{m}_2)^2} \\
+ V(\bm{r}) - E\biggr] \psi(\bm{r},\bm{R})=0,
\label{eq:schrnonsep}
\end{multline}
where
\begin{equation}
\mu = \frac{m_1 m_2}{m_1 + m_2}
\end{equation}
is a diagonal reduced-mass matrix.
The system of Schr\"odinger equations is exactly separable if the matrix multiplying $\bm{p}\cdot\bm{P}$ is zero, which requires
\begin{equation}
\frac{\overline{m}_1}{m_1} = \frac{\overline{m}_2}{m_2}.
\label{eq:sepcond}
\end{equation}
In this case, Equation~\eqref{eq:schrnonsep} becomes
\begin{equation}
\biggl[\frac{p^2}{2 \mu} + \frac{P^2}{2(m_1 + m_2)} + V(\bm{r}) - E\biggr] \psi(\bm{r},\bm{R})=0.
\label{eq:schrsep}
\end{equation}
Then, separating $\bm{R}$ and its conjugate momentum $\bm{P}$ from Equation~\eqref{eq:schrsep} produces a differential equation in the single vector variable $\bm{r}$,
\begin{equation}
\biggl[\frac{p^2}{2 \mu} + V(\bm{r}) - E\biggr] \psi(\bm{r})=0,
\label{eq:schr1bod}
\end{equation}
where $E$ now represents the energy in the center-of-mass frame where $\bm{P}=0$.

Note that the exact separability condition in Equation~\eqref{eq:sepcond} can be satisfied if and only if the two mass matrices are linearly dependent, for it requires $m_1 \propto m_2$.
If $m_1$ and $m_2$ are linearly independent, Equation~\eqref{eq:sepcond} cannot be satisfied but Equation~\eqref{eq:schrnonsep} may still be approximately separable if
\begin{equation}
\biggl\lVert\frac{\overline{m}_1}{m_1} - \frac{\overline{m}_2}{m_2}\biggr\rVert \ll 1,
\end{equation}
where the double vertical lines denote the matrix norm.
That is typically the case for a system of Schr\"odinger equations describing a nonrelativistic scattering process.
The mass difference between the initial and final states coincides with the amount of kinetic energy exchanged during the collision.
The nonrelativistic limit holds under the assumption that the kinetic energy of each particle is much smaller than its mass.
Thus, the mass differences $\lvert m_{1,a} - m_{1,b} \rvert$ and $\lvert m_{2,a} - m_{2,b} \rvert$ with $a\neq b$ can be assumed to be much smaller than the masses $m_{1,a}$ and $m_{2,a}$, respectively.
We can therefore write
\begin{equation}
\frac{\overline{m}_1}{m_1} = I + \Delta_1 \qquad \text{and} \qquad \frac{\overline{m}_2}{m_2} = I + \Delta_2,
\label{m1m2approx}
\end{equation}
with $I$ for the identity matrix, and we see that the matrices $\Delta_1=(\overline{m}_1 - m_1)/m_1$ and $\Delta_2=(\overline{m}_2 - m_2)/m_2$ are ``small'', that is, $\lVert \Delta_1 \rVert \ll 1$ and $\lVert \Delta_2 \rVert \ll 1$.
It then follows from Equation~\eqref{m1m2approx} and the Cauchy-Schwartz inequality that
\begin{equation}
\biggl\lVert\frac{\overline{m}_1}{m_1} - \frac{\overline{m}_2}{m_2} \biggr\rVert
= \lVert \Delta_1 - \Delta_2 \rVert \leq \lVert \Delta_1 \rVert + \lVert \Delta_2 \rVert \ll 1.
\end{equation}

We now wish to further reduce Equation~\eqref{eq:schr1bod} to a system of differential equations in the distance $r = \lvert \bm{r} \rvert$.
The Schr\"odinger equation in the channel $a$ reads
\begin{equation}
\sum_{b=1}^\mathcal{N}\biggl[\delta_{ab}\frac{p^2}{2 \mu_a} + V_{ab}(\bm{r}) - \delta_{ab} E\biggr] \psi_b(\bm{r})=0,
\label{eq:schrexplicit}
\end{equation}
where $\mu_a = m_{1,a} m_{2,a}/(m_{1,a} + m_{2,a})$ is the reduced mass in the channel $a$, $V_{ab}(\bm{r})$ is the potential coupling the channel $b$ to the channel $a$, and $\psi_a(\bm{r})$ is the wavefunction in the channel $a$.
Recall that $\psi_a(\bm{r})$ contains the spin states of the particles in the channel $a$, and let $J_{1,a}$ and $J_{2,a}$ be their spins.%
\footnote{%
Here we refer to the total angular momentum of each individual particle as its ``spin''. We reserve the term ``total angular momentum'' for the entire two-particle system.%
}
We can therefore expand the wavefunction $\psi_a(\bm{r})$ as
\begin{equation}
\psi_a(\bm{r}) = \sum_{s=\lvert J_{1,a} - J_{2,a}\rvert}^{J_{1,a} + J_{2,a}} \sum_{l=0}^\infty \sum_{\sigma=-s}^s \sum_{m=-l}^l \psi_{a,s,l}^{\sigma,m}(r) Y_l^m(\theta,\phi) \xi_s^\sigma,
\label{eq:expand}
\end{equation}
where $\psi_{a,s,l}^{\sigma,m}(r)$ are radial wavefunctions, $Y_l^m(\theta,\phi)$ are spherical harmonics, and $\xi_s^\sigma$ are irreducible representations from the decomposition of the direct product of the spin states $\chi_{J_{1,a}}^{M_{1,a}}$ and $\chi_{J_{2,a}}^{M_{2,a}}$ of the individual particles,
\begin{equation}
\xi_s^\sigma = \sum_{M_{1,a}=-J_{1,a}}^{J_{1,a}} \sum_{M_{2,a}=-J_{2,a}}^{J_{2,a}} \langle J_{1,a} J_{2,a} M_{1,a} M_{2,a} \vert J_{1,a} J_{2,a} s \, \sigma\rangle \chi_{J_{1,a}}^{M_{1,a}} \chi_{J_{2,a}}^{M_{2,a}},
\label{eq:spindeomp}
\end{equation}
where the term with angle brackets is a Clebsch-Gordan coefficient.
The radial wavefunctions in Equation~\eqref{eq:expand} are labeled by quantum numbers $\sigma$ and $m$ for the third components of the total spin and orbital angular momentum, respectively.
It is more convenient to rewrite Equation~\eqref{eq:expand} as an expansion in radial wavefunctions labeled by quantum numbers $J$ and $M$ for the total angular momentum,
\begin{equation}
\psi_a(\bm{r}) =  \sum_{s=\lvert J_{1,a} - J_{2,a}\rvert}^{J_{1,a} + J_{2,a}} \sum_{l=0}^\infty \sum_{J=\lvert s - l \rvert}^{s + l} \sum_{M=-J}^J \psi_{a,s,l}^{J,M}(r) \sum_{\sigma,m} \langle s l \sigma m \vert s l J M \rangle Y_l^m(\theta,\phi) \xi_s^\sigma.
\label{eq:wavefunc}
\end{equation}

Finally, one can derive the system of radial Schr\"odinger equations by substituting Equation \eqref{eq:wavefunc} into \eqref{eq:schrexplicit}, multiplying on the left by
\[
\sum_{\sigma^\prime=-s}^s \sum_{m^\prime=-l}^l \langle s^\prime l^\prime \sigma^\prime m^\prime \vert s^\prime l^\prime J^\prime M^\prime \rangle Y_{l^\prime}^{m^\prime}(\theta,\phi)^\ast \xi_{s^\prime}^{\sigma^\prime\dag},
\]
integrating over the angles $\theta$ and $\phi$, and imposing conservation of angular momentum.
This produces infinitely many systems of radial Schr\"odinger equations labeled by conserved quantum numbers $J$ and $M$ for the total angular momentum and its third component.
Systems labeled by different total-angular-momentum quantum numbers are decoupled from each other.
Using matrix notation, each system can be written concisely as
\begin{equation}
\biggl[\frac{1}{2\mu}\biggl(- \frac{\mathrm{d}^2\hphantom{r}}{\mathrm{d}r^2} - \frac{2}{r}\frac{\mathrm{d}\hphantom{r}}{\mathrm{d}r} + \frac{L (L+1)}{r^2}\biggr) + V^J(r) - E \biggr] \psi^{J,M}(r)=0,
\label{eq:schrad}
\end{equation}
where $L$ is a diagonal orbital-angular-momentum matrix, $V^J(r)$ is a radial potential matrix, and $\psi^{J,M}(r)$ is a vector wavefunction.
Note that the vector wavefunction depends on both $J$ and $M$, but the potential matrix depends only on $J$.
The radial Schr\"odinger equations in this system are labeled by triplets $(a,l,s)$ that identify spin-orbital channels of the scattering process.
Since $s$ and $l$ are constrained by the triangle relations $\lvert J_{1,a} - J_{2,a}\rvert\leq s \leq J_{1,a} + J_{2,a}$ and $\lvert J - s\rvert \leq l \leq J + s$, the system in Equation~\eqref{eq:schrad} contains only a finite number of radial Schr\"odinger equations.
To simplify the notation, we specify each orbital channel by a single integer $i$ instead of the three labels $a$, $s$, and $l$.
Then the radial potential matrix elements can be expressed as
\begin{multline}
V_{ij}^J(r) = \sum_{\sigma_i=-s_i}^{s_i} \sum_{m_i=-l_i}^{l_i} \sum_{\sigma_j=-s_j}^{s_j} \sum_{m_j=-l_j}^{l_j} \langle s_i l_i \sigma_i m_i \vert s_i l_i J M \rangle \langle s_j l_j \sigma_j m_j \vert s_j l_j J M \rangle \\
\times\int \mathrm{d}\Omega \, Y_{l_i}^{m_i}(\theta,\phi)^\ast Y_{l_j}^{m_j}(\theta,\phi) \xi_{s_i}^{\sigma_i\dag} V_{a_i b_j}(\bm{r}) \xi_{s_j}^{\sigma_j},
\label{eq:pot}
\end{multline}
where $i \to (a_i,s_i,l_i)$ and $j \to (b_j,s_j,l_j)$. Note that each potential $V_{a_i b_j}(\bm{r})$ is an operator mapping spin vectors in the channel $b_j$ to spin vectors in the channel $a_i$ and that the total spin states $\xi_{s_i}^{\sigma_i}$ and $\xi_{s_j}^{\sigma_j}$ can be expanded in terms of the spin states of the individual particles in the channels $a_i$ and $b_j$ using Equation~\eqref{eq:spindeomp}.
Then we see that the right-hand side of Equation~\eqref{eq:pot} does not depend on $M$ by virtue of the Wigner-Eckart theorem.

There is a somewhat common misunderstanding that the reduction of a three-dimensional Schr\"odinger equation to a radial form can be performed only for a spherically-symmetric potential that does not depend on the angles $\theta$ and $\phi$.
However, in our derivation of Equations~\eqref{eq:schrad} and \eqref{eq:pot} we have made no assumption on the potential $V(\bm{r})$ other than it conserves angular momentum.
In the trivial case where the potential is independent of spin, then conservation of angular momentum does indeed require that it is also spherically symmetric.
If the potential does depend on spin, then it can also depend on the angles $\theta$ and $\phi$ but Equations~\eqref{eq:schrad} and \eqref{eq:pot} would still apply.

The system in Equation~\eqref{eq:schrad} describes a scattering process taking place in a partial wave labeled by quantum numbers $J$ and $M$.
Because of conservation of angular momentum, each partial wave is decoupled from the others.
SPARSE solves the scattering problem only for one individual partial wave, leaving the task of summing partial waves to the user.
Hence, from now on we will drop all superscripts for the conserved quantum numbers $J$ and $M$.

To conclude this section, we simplify the derivatives inside Equation~\eqref{eq:schrad} as follows. Let us observe that
\begin{equation}
\biggl(\frac{\mathrm{d}^2\hphantom{r}}{\mathrm{d}r^2} + \frac{2}{r}\frac{\mathrm{d}\hphantom{r}}{\mathrm{d}r}\biggr)\psi(r) = \frac{1}{r}\frac{\mathrm{d}^2 \hphantom{r}}{\mathrm{d}r^2} u(r),
\label{eq:redwfdiff}
\end{equation}
where we have introduced the reduced radial wavefunction
\begin{equation}
u(r) = r \psi(r).
\end{equation}
Substituting Equation~\eqref{eq:redwfdiff} into \eqref{eq:schrad} and multiplying both sides by $r$ yields
\begin{equation}
\biggl[-\frac{1}{2\mu} \frac{\mathrm{d}^2\hphantom{r}}{\mathrm{d}r^2} + \frac{L (L+1)}{2\mu r^2} + V(r) - E \biggr] u(r)=0.
\label{eq:schr}
\end{equation}
Equation~\eqref{eq:schr} is the system of radial Schr\"odinger equations solved by the SPARSE algorithm.
It has $N$ spin-orbital channels labeled by an integer $i=1,2,\dots,N$.
The channel $i$ has spin $s_i$ and orbital angular momentum $l_i$.
The reduced-mass matrix elements are $\mu_{ij}=\delta_{ij}\mu_i$.
The orbital-angular-momentum matrix elements are $L_{ij} = \delta_{ij} l_i$.
The radial potential matrix elements $V_{ij}(r)$ are given in Equation~\eqref{eq:pot}.

\section{Numerical solution}
\label{sec:wavefunc}

To solve Equation~\eqref{eq:schr} numerically, we truncate and discretize the coordinate space for the distance $r$ from $\mathbb{R}_0^+$ to a finite grid of equally-spaced points $\{r_n\}_{n=0}^{M+1}$, where
\begin{equation}
r_n=n d
\end{equation}
with $d$ a small discretization step.
Our discretized coordinate space begins at the origin $r_0=0$, has $M$ points in its interior that we refer to as \emph{nodes}, and ends at a finite but otherwise arbitrarily large radius $r_{M+1}=R$.
The numerical reduced radial wavefunction in the channel $i$ is $u_i(r_n)$.
The numerical potential matrix element coupling the channel $j$ to the channel $i$ is $V_{ij}(r_n)$.
The numerical second derivative of the reduced radial wavefunction is evaluated using the leading-order finite-difference method,
\begin{equation}
\frac{\mathrm{d}^2\hphantom{r}}{\mathrm{d}r^2} u(r_n) = \frac{u(r_{n - 1}) - 2 u(r_n) + u(r_{n + 1})}{d^2} + \mathcal{O}(d^4).
\label{eq:secder}
\end{equation}
For small enough discretization steps $d$, the $\mathcal{O}(d^4)$ corrections can be neglected.

To close the system of equations, we fix the values of the reduced radial wavefunctions at the boundaries, $u_i(0)$ and $u_i(R)$. This is referred to as imposing Dirichlet boundary conditions. We can choose the physical boundary conditions using the known analytical behavior of the solutions to Equation~\eqref{eq:schr} for $r\to0$ and $r\to\infty$.
Since a physical solution $u_i(r)$ goes to zero as $r^{l_i+1}$ for $r\to 0$, we impose
\begin{equation}
u_i(0)=0
\label{eq:boundmin}
\end{equation}
for every $i$. The boundary condition at large $r$ is more complicated.
For physical scattering problems, one has
\begin{equation}
\lim_{r\to\infty} V_{ij}(r) = \delta_{ij} V_i^\infty
\end{equation}
where each $V_i^\infty$ is either a definite number or positive infinity.
We identify $V_i^\infty$ with the energy threshold where the scattering channel $i$ opens up.
The boundary condition for $u_i(R)$ depends on whether the energy $E$ in the Schr\"odinger equation is larger or smaller than the threshold $V_i^\infty$.
If $E < V_i^\infty$, then a physical solution goes to zero for $r\to\infty$ and we impose $u_i(R)=0$.
If otherwise $E \geq V_i^\infty$, a physical solution keeps oscillating indefinitely for $r\to\infty$.
It is however a specific, finite number at the maximum distance $R$ in our truncated coordinate space.
So, we impose $u_i(R)=U_i$, where $U_i$ is a real constant.
We can conveniently impose the boundary condition for both cases $ E < V_i^\infty$ and $E \geq V_i^\infty$ at once in the form
\begin{equation}
u_i(R)= \theta(E - V_i^\infty) U_i,
\label{eq:boundmax}
\end{equation}
where
\begin{equation}
\theta(x) =
\begin{cases}
0 & \text{if $x < 0$,} \\
1 & \text{if $x \geq 0$,}\\
\end{cases}
\end{equation}
is the Heaviside step function.

Using Equations~\eqref{eq:secder}, \eqref{eq:boundmin}, and \eqref{eq:boundmax}, we reduce Equation~\eqref{eq:schr} down to a system of numerical Schr\"odinger equations,
\begin{equation}
\sum_{j=1}^N \sum_{m=1}^M\bigl[H_{ij,nm} - \delta_{ij}\delta_{nm} E\bigr] u_j(r_m) \\
= \delta_{nM} \theta(E - V_i^\infty) \frac{U_i}{2\mu_i d^2},
\label{eq:schrnum}
\end{equation}
where $H_{ij,nm}$ are the numerical Hamiltonian matrix elements
\begin{equation}
H_{ij,nm} =- \delta_{ij}\frac{\delta_{(n-1)m} - 2 \delta_{nm} + \delta_{(n+1)m}}{2\mu_i d^2} + \delta_{ij}\delta_{nm}\frac{l_i (l_i+1)}{2\mu_i r_n^2} + \delta_{nm} V_{ij}(r_n)
\label{eq:hamiltonian}
\end{equation}
with $i,j=1,\dots,N$ denoting channels and $n,m=1,\dots,M$ denoting nodes.
The numerical Hamiltonian is a matrix with $N^2 M^2$ elements, but only a small fraction of them are different from zero.
The ratio of nonzero elements over the total scales like $1/M$, with $M$ being a very large number for a large maximum distance $R$ and a small discretization step $d$.
The numerical Hamiltonian in Equation~\eqref{eq:hamiltonian} is therefore a sparse matrix.

To input the Schr\"odinger equation into SPARSE, the user has to provide the following information:
\begin{itemize}
\item The orbital angular momentum $l_i$, reduced mass $\mu_i$, and threshold $V_i^\infty$ for each channel $i$.
\item The positions of the nodes $r_n$ and the numerical potential matrix elements $V_{ij}(r_n).$
\end{itemize}
SPARSE reads this information from two comma-separated-values (CSV) files named \verb!channels.csv! and \verb!potential.csv!, which must be placed within the working directory.
Note that all the required spin information is encoded in the potential matrix elements through Equation~\eqref{eq:pot}.
The remaining input, the total energy $E$, will be specified by the user as an argument of the functions defined in the SPARSE module.

The file \verb!channels.csv! must have one line for the header followed by $N$ lines for the values.
Each line must contain at least three entries.
The header must contain at least the following strings: \verb!l!, \verb!mu!, and \verb!threshold!.
The following $N$ lines must contain at least the corresponding values for the channels $i=1,\dots,N$.
An entry under \verb!l! is an integer indicating $l_i$.
An entry under \verb!mu! is a decimal number indicating $\mu_i$.
An entry under \verb!threshold! is a decimal number indicating $V_i^\infty$ or \verb!inf! if $V_i^\infty=\infty$.
The ordering of the columns is irrelevant, but it must obviously be consistent between rows.
Below is the typical structure of a minimal \verb!channels.csv! file.
\begin{verbatim}
l,mu,threshold
l1,mu1,T1
l2,mu2,T2
...
lN,muN,TN
\end{verbatim}
For the user's convenience, additional information about the channels may be specified by adding more columns to the file \verb!channels.csv!.
For instance, the user may want to include a column named \verb!s! indicating the spin $s_i$ for each channel $i$, or a column named \verb!channel! providing a label for each channel that is more informative than just the integer $i$.
Such additional columns will be read by SPARSE but will not play any role in the calculation.

The file \verb!potential.csv! must have no header and $M$ lines for the values.
Each line must contain exactly $N^2 + 1$ entries.
The first entry is the node position $r_n$.
The following $N^2$ entries are the numerical values of the flattened potential matrix, $V_{ij}(r_n)$.
SPARSE assumes that the potential matrix is flattened in row-major (C-style) order.%
\footnote{%
This fact is irrelevant if the potential is a symmetric matrix.%
}
Below is the typical structure of the \verb!potential.csv! file.
\begin{verbatim}
r1,V11(r1),V12(r1),...,VNN(r1)
r2,V11(r1),V12(r2),...,VNN(r2)
...
rM,V11(rM),V12(rM),...,VNN(rM)
\end{verbatim}

SPARSE uses natural units, in which $\hbar=c=1$ and therefore $[\text{Time}]=[\text{Length}]$, $[\text{Mass}]=[\text{Energy}]$, and $[\text{Length}]=[\text{Energy}]^{-1}$.
There is only one independent dimension, typically chosen between $[\text{Length}]$ in units of fm (femtometers) and $[\text{Energy}]$ in units of eV (electron Volts).
The conversion between length-based and energy-based units is easily achieved by means of the formula $\hbar c = 1 = 197.327$ MeV fm.
The SPARSE algorithm is agnostic to the user's choice between length-based or energy-based units, but it assumes that the dimensions of the inputs are consistent with $[r_n]=[\text{Length}]$ and $[V_{ij}(r_n)]=[\mu_i]=[V_i^\infty]=[\text{Energy}]$.

SPARSE reads the values from the two CSV files \verb!channels.csv! and \verb!potential.csv! and stores them into two pandas \verb!DataFrame! objects using the function \verb!read_csv! from the pandas library \cite{pandas}.
From them, SPARSE calculates the Hamiltonian matrix elements $H_{ij,nm}$ from Equation~\eqref{eq:hamiltonian} and writes them into a 2-dimensional NumPy \verb!array! object \cite{NumPy} with elements $H_{\mu\nu}$.
This reshaping amounts to applying an invertible map $(i,n)\mapsto\mu$ from pairs of indices $i=1,\dots,N$ and $n=1,\dots,M$ to a single index $\mu=1,\dots,NM$.
SPARSE reads the matrix elements from Equation~\eqref{eq:hamiltonian} and writes them in the NumPy \verb!array! following a C-style index order, which corresponds to setting
\begin{equation}
\mu = i + N (n - 1).
\label{eq:reshape}
\end{equation}
This choice of index order has the advantage that all the nonzero $H_{\mu\nu}$ end up as close as possible to the main diagonal.
From Equation~\eqref{eq:hamiltonian}, we see that the only nonzero $H_{ij,nm}$ are those with $i=j$ and $\lvert n - m \rvert \leq 1$ or those with $n=m$ and $\lvert i - j \rvert \leq N-1$.
Then it follows from Equation~\eqref{eq:reshape} that the only nonzero $H_{\mu\nu}$ are those with
\begin{equation}
\lvert\mu - \nu\rvert \leq \lvert i - j \rvert + N \lvert n - m \rvert \leq N.
\end{equation}
Therefore the $H_{\mu\nu}$ form a banded matrix where all the nonzero elements are contained within the main diagonal, the first $N$ upper diagonals, and the first $N$ lower diagonals.

SPARSE writes the Hamiltonian in the matrix diagonal ordered form, which is a 2-dimensional array consisting of only the $2N+1$ nonzero diagonals stacked from top to bottom.
The upper and lower diagonals are padded so that they all have the same number of elements $N M$ as the main diagonal.
The matrix diagonal ordered form has only $(2N + 1) \times (NM)$ elements, and therefore it takes significantly less memory than the full $(NM) \times (NM)$ numerical Hamiltonian matrix with its many zeroes.
Consider for example a system with $N=4$ channels and $M=10^6$ nodes using 64-bit floating-point numbers.
The matrix diagonal ordered form would take 288 megabytes of memory, as opposed to a staggering 128 terabytes for the full matrix.
If not for this memory optimization, it would be all but impossible to accurately solve a large system of Schr\"odinger equations using the finite-difference method.

We therefore redefine the indices of Equation~\eqref{eq:schrnum} as $(i,n)\to\mu$ and $(j,m)\to\nu$ using Equation~\eqref{eq:reshape}, which reshapes Equation~\eqref{eq:schrnum} into an ordinary nonhomogeneous linear system,
\begin{equation}
\sum_{\nu=1}^{NM} (H_{\mu\nu} - \delta_{\mu\nu} E) u_\nu = B_\mu(E),
\label{eq:linsyst}
\end{equation}
where the constant vector is
\begin{equation}
B_\mu(E) = \delta_{nM} \theta(E - V_i^\infty) \frac{U_i}{2\mu_i d^2}.
\label{eq:veconst}
\end{equation}

Note that if $E < V_i^\infty$ for all $i$, which corresponds to a bound state, then $B_\mu(E)=0$ for all $\mu$ and Equation~\eqref{eq:linsyst} becomes an eigenvalue problem for the Hamiltonian matrix. Even though SPARSE is primarily designed for calculating scattering states, it also contains a handy bound-state calculator which solves Equation~\eqref{eq:linsyst} with $B_\mu(E)=0$ using the function \texttt{eigsh} from the sparse linear algebra module of SciPy \cite{SciPy}.

From here on we shall assume, without loss of generality, that the channels $i$ are ordered in increasing threshold value.
It is worth noting that the SPARSE algorithm does not require the input channels to be sorted in any particular way.
The assumption $V_i^\infty \geq V_j^\infty$ for any $i \geq j$ is only made to simplify the notation in the rest of the paper.

For a given energy $E$, the possible values of the constant in Equation~\eqref{eq:veconst} constitute a vector space of dimension $O$, where $O$ is the number of open channels $i$ whose threshold is less than $E$, $V_i^\infty \leq E$ for $i=1,\dots,O$.
We can construct a basis for this vector space by substituting $U_i$ inside Equation~\eqref{eq:veconst} with a two-index term $U_{ij}$, where the first index $i=1,\dots,N$ labels the channels and the second index $j=1,\dots,O$ labels the vectors in the basis.
Note that, for the basis to be complete, the $U_{ij}$ with $i\leq O$ must form the elements of an invertible $O \times O$ matrix.
Substituting $U_i\to U_{ij}$ in Equation~\eqref{eq:veconst}, we obtain a constant term with two indices, $B_{\mu j}(E)$, which we can treat as the elements of a $(NM)\times O$ matrix $B$, dropping its dependence on $E$.
We also define the $(NM)\times(NM)$ matrix
\begin{equation}
A = H - E I.
\label{eq:amatrix}
\end{equation}
with $I$ the identity matrix.
In practice, $A$ is a banded matrix which SPARSE calculates by applying an offset of $-E$ to the main diagonal of the numerical Hamiltonian in the matrix diagonal ordered form.
With these definitions, we rewrite Equation~\eqref{eq:linsyst} as a matrix equation
\begin{equation}
A u = B,
\label{eq:matsyst}
\end{equation}
where $u$ is a $(NM)\times O$ matrix whose columns form a complete set of linearly independent solutions to Equation~\eqref{eq:linsyst} with the same energy $E$.

For any energy $E$, SPARSE constructs the matrices $A$ and $B$ using Equations~\eqref{eq:veconst}-\eqref{eq:amatrix} and solves Equation~\eqref{eq:matsyst} for the matrix $u$ using LAPACK general banded LU factorization routines through the \verb!linalg.solve_banded! function from SciPy \cite{SciPy}. Then, SPARSE reshapes the matrix $u$ into a set of numerical reduced wavefunctions, $u_{\mu j} \to u_{ij}(r_n)$, by inverting the index mapping $(i,n)\mapsto \mu$ defined in Equation~\eqref{eq:reshape}.

\section{Calculating the \texorpdfstring{$K$}{K}-matrix}
\label{sec:kmatrix}

We proceed to calculate the reactance matrix $K$ for the scattering process from the numerical wavefunctions $u_{ij}(r_n)$. We will disregard from now on wavefunction components for confining channels $i$ corresponding to $V_i^\infty=\infty$, which are irrelevant to the scattering states at large $r$.

Let us first define the numerical range of the potential $r_V$ as the maximum value of $r$ for which any diagonal potential $V_{ii}(r)$ is significantly different from its associated threshold%
\footnote{Potentials that diverge at $r\to\infty$ do not have a threshold and, by definition, they have an infinite radius. We neglect them since they correspond to confining channels.}
or any off-diagonal potential $V_{ij}(r)$ with $i\neq j$ is significantly different from zero.
For $r > r_V$, the system in Equation~\eqref{eq:schr} reduces to a set of decoupled Schr\"odinger equations for free particles,
\begin{equation}
\biggl[-\frac{1}{2\mu_{i}} \frac{\mathrm{d}^2\hphantom{r}}{\mathrm{d}r^2} + \frac{l_i (l_i+1)}{2\mu r^2} + V_i^\infty - E \biggr] u_i(r) = 0.
\label{eq:schrfree}
\end{equation}
For a closed channel $i=O+1,\dots,N$ with $V_i^\infty > E$, a physical solution to Equation~\eqref{eq:schrfree} must approach zero at large distances.
Therefore, we further neglect wavefunction components for the closed channels.
For an open channel $i=1,\dots,O$ with $V_i^\infty \leq E$, Equation~\eqref{eq:schrfree} has two linearly independent physical solutions. Using real boundary conditions and wavefunction normalization by energy (see Reference~\cite{Mor07} for alternative conventions), the two solutions are
\begin{equation}
\sqrt{\frac{2 \mu_i}{\pi p_i}} S_{l_i}(p_i r) \qquad \text{and} \qquad \sqrt{\frac{2 \mu_i}{\pi p_i}} C_{l_i}(p_i r),
\label{eq:freesol}
\end{equation}
where $p_i$ is the scattering momentum in the open channel $i$,
\begin{equation}
p_i = \sqrt{2 \mu_i (E - V_i^\infty)},
\end{equation}
and $S_l(x)$ and $C_l(x)$ are the Riccati-Bessel functions
\begin{align}
S_l(x) &= x j_l(x), \\
C_l(x) &= -x y_l(x),
\end{align}
with $j_l(x)$ and $y_l(x)$ the spherical Bessel functions of the first and second kind, respectively.
The asymptotic solution of any Schr\"odinger equation that coincides with Equation~\eqref{eq:schrfree} at large $r$ can be written as a linear combination of the solutions in Equation~\eqref{eq:freesol}.
In particular, the asymptotic wavefunctions for a complete set of scattering states with energy $E$ can be expressed as
\begin{equation}
u_{ij}(r) \simeq \sqrt{\frac{2 \mu_i}{\pi p_i}} \bigl(S_{l_i}(p_i r) + K_{ij} C_{l_i}(p_i r) \bigr),
\label{eq:asymsol}
\end{equation}
where the relation ``$\simeq$'' stands for equality at large $r$, the first index $i=1,\dots,O$ labels open channels, the second index $j=1,\dots,O$ labels linearly independent solutions with the same energy, and $K_{ij}$ are $K$-matrix elements (see Section~11.1.2 of Reference~\cite{New13}).

Next, we compare the numerical wavefunctions $u_{ij}(r_n)$ to the analytic asymptotic wavefunctions in Equation~\eqref{eq:asymsol} to extract the $K$-matrix elements.
The numerical wavefunctions calculated from Equation~\eqref{eq:matsyst} generally do not correspond to the asymptotic scattering states in Equation~\eqref{eq:asymsol}.
Instead, the numerical wavefunctions at large $r$ correspond to a more general linear combination of Riccati-Bessel functions,
\begin{equation}
u_{ij}(r) \simeq \sqrt{\frac{2 \mu_i}{\pi p_i}} \bigl( X_{ij} S_{l_i}(p_i r) + Y_{ij} C_{l_i}(p_i r) \bigr),
\label{eq:numsol}
\end{equation}
where $X_{ij}$ and $Y_{ij}$ are real constants that can be regarded as the elements of two $O\times O$ matrices $X$ and $Y$.
Comparing Equations~\eqref{eq:asymsol} and \eqref{eq:numsol}, we see that the $K$-matrix can be calculated from the matrices $X$ and $Y$ as
\begin{equation}
K = Y X^{-1}.
\label{eq:kmatrix}
\end{equation}
The matrices $X$ and $Y$ depend on the choice of the matrix constant $B$ in Equation~\eqref{eq:matsyst}, but the $K$-matrix does not.
If instead of $B$ we had chosen any other admissible $B^\prime$, then instead of $X$ and $Y$ we would have obtained $X^\prime = X W$ and $Y^\prime=Y W$ with $W=B^{-1}B^\prime$ an invertible $O\times O$ matrix. However, the $K$-matrix from Equation~\eqref{eq:kmatrix} would have been the same.

Finally, let us show how to calculate the matrices $X$ and $Y$ explicitly from the numerical wavefunctions $u_{ij}(r_n)$.
We fit the numerical wavefunctions $u_{ij}(r_n)$ to the analytical form in Equation~\eqref{eq:numsol} by minimizing the loss function
\begin{equation}
L(X,Y) = \sum_{n=n_V}^N \sum_{i=1}^M \sum_{j=1}^O \biggl\lvert \sqrt{\frac{2 \mu_i}{\pi p_i}} \bigl( X_{ij} S_{l_i}(p_i r_n) + Y_{ij} C_{l_i}(p_i r_n)\bigr) - u_{ij}(r_n) \biggr\rvert^2,
\end{equation}
where $n_V$ is the index corresponding to the smallest $r_n$ that exceeds the radius $r_V$ of the potential, $n_V = \min \{ n \mid r_n > r_V\}$.
The least-square fits of $X$ and $Y$ are obtained by solving
\begin{equation}
\frac{\partial L}{\partial X} = \frac{\partial L}{\partial Y} = 0.
\end{equation}
The above equation produces a set of independent linear systems, one for each $i$ and each $j$,
\begin{equation}
\begin{pmatrix}
\alpha_i & \gamma_i \\
\gamma_i & \beta_i
\end{pmatrix}
\begin{pmatrix}
X_{ij} \\
Y_{ij}
\end{pmatrix}
=
\begin{pmatrix}
\mathcal{S}_{ij} \\
\mathcal{C}_{ij}
\end{pmatrix},
\label{eq:abcscsystem}
\end{equation}
where we have introduced the shorthand
\begin{equation}
\alpha_i = \sum_{n=n_V}^N S_{l_i}(p_i r_n)^2, \quad
\beta_i = \sum_{n=n_V}^N C_{l_i}(p_i r_n)^2, \quad
\gamma_i = \sum_{n=n_V}^N S_{l_i}(p_i r_n) C_{l_i}(p_i r_n),
\label{eq:abc}
\end{equation}
and
\begin{align}
\mathcal{S}_{ij} = \sqrt{\frac{\pi p_i}{2 \mu_i}} \sum_{n=n_V}^N u_{ij}(r_n) S_{l_i}(p_i r_n), \quad
\mathcal{C}_{ij} = \sqrt{\frac{\pi p_i}{2 \mu_i}} \sum_{n=n_V}^N u_{ij}(r_n) C_{l_i}(p_i r_n) .
\label{eq:sc}
\end{align}
The resulting expressions for the matrix elements of $X$ and $Y$ are
\begin{equation}
X_{ij} = \frac{\beta_i \mathcal{S}_{ij} - \gamma_i \mathcal{C}_{ij}}{\alpha_i \beta_i - \gamma_i^2}
\label{eq:x}
\end{equation}
and
\begin{equation}
Y_{ij} = \frac{\alpha_i \mathcal{C}_{ij} - \gamma_i \mathcal{S}_{ij}}{\alpha_i \beta_i - \gamma_i^2},
\label{eq:y}
\end{equation}
respectively.
SPARSE evaluates the right-hand sides of Equations~\eqref{eq:abc} and \eqref{eq:sc} using the numerical solutions of the Schr\"odinger equation and the spherical Bessel functions from SciPy \cite{SciPy}.
It is important to note that the denominator on the right-hand side of both Equations~\eqref{eq:x} and \eqref{eq:y} vanishes in the limit $p_i\to 0$.
Concretely, the right-hand sides of Equations~\eqref{eq:x} and \eqref{eq:y} become numerically unstable for $p_i \ll ((R - r_V)^{-1}$.
One should therefore choose a suitably large maximum radius $R$ to calculate scattering in an energy region very close to a threshold.

\section{Scattering poles and amplitudes}
\label{sec:scattering}

We define the transition matrix $T$ in terms of the scattering matrix $S$ as
\begin{equation}
S = I + 2 i T.
\label{eq:smatrix}
\end{equation}
Within this convention, a $T$-matrix element $T_{ij}$ corresponds to the scattering amplitude $f_{i\leftarrow j}$ from an initial channel $j$ to a final channel $i$ times the relative momentum in the final channel,
\begin{equation}
T_{ij} = p_i f_{i\leftarrow j}.
\label{eq:tfmatrix}
\end{equation}
Scattering resonances are associated with poles of the $T$-matrix for energies on an unphysical Riemann sheet.
Most commonly, resonances correspond to poles located on the lower half of the unphysical sheet that connects to the real axis of the physical sheet.
The partial-wave cross-section, $\sigma_{i\leftarrow j}$, is proportional to the square modulus of the scattering amplitude,
\begin{equation}
\sigma_{i\leftarrow j} = \frac{4\pi(2J+1)}{(2 J_{1,a_j} + 1) (2 J_{2,a_j} + 1)} \lvert f_{i\leftarrow j} \rvert^2,
\end{equation}
where $J$ is the total angular momentum and $J_{1,a_j}$ and $J_{2,a_j}$ are the spins of the interacting particles in the initial channel $a_j$.
The total unpolarized cross section can be obtained by summing over the spins $s_i$, $s_j$, the orbital angular momenta $l_i$, $l_j$, and the total angular momentum $J$.

The reactance matrix $K$ is defined as the Cayley transform of the $S$-matrix,
\begin{equation}
K = i\frac{I - S}{I + S},
\label{eq:ksmatrix}
\end{equation}
which can be inverted to give
\begin{equation}
S = \frac{I + i K}{I - i K}
\label{eq:skmatrix}
\end{equation}
(see Section~14-e of Reference~\cite{Tay83}).
We see from Equations~\eqref{eq:ksmatrix} and \eqref{eq:skmatrix} that unitarity ($S^\dag=S^{-1}$) and time-reversal symmetry ($S^\mathsf{T} = S$) require that the $K$-matrix is real and symmetric.
The numerical $K$-matrix is real by definition.
It is only approximately symmetric within numerical errors.
Let us call this approximately symmetric matrix $\tilde{K}$.
SPARSE calculates the degree of asymmetry as
\begin{equation}
\max_{ij}\Biggl\{\frac{\bigl\lvert \tilde{K}_{ij} - \tilde{K}_{ji}\bigr\rvert}{\bigl\lvert \tilde{K}_{ij} + \tilde{K}_{ji}\bigr\rvert} \Biggr\}
\end{equation}
and warns the user if it is larger than a small tolerance. Then, SPARSE calculates the exactly symmetric $K$-matrix as
\begin{equation}
K = \frac{\tilde{K} + \tilde{K}^\mathsf{T}}{2}.
\end{equation}
Finally, the $T$-matrix can be calculated from the numerical $K$-matrix elements by inserting Equation~\eqref{eq:skmatrix} into \eqref{eq:smatrix}, which gives
\begin{equation}
T = \frac{K}{I - i K}.
\label{eq:tkmatrix}
\end{equation}

SPARSE calculates scattering poles by interpolating between numerical $T$-matrix elements evaluated at a discrete set of real energies.%
\footnote{SPARSE can also calculate so-called \emph{nominal} scattering poles by interpolating between numerical $K$-matrix elements evaluated at a discrete set of real energies.}
The interpolation is performed using the AAA algorithm for rational polynomial approximation \cite{AAA}.
For a function $F$ evaluated at a discret set of energies, the interpolating rational function is represented in the barycentric form,
\begin{equation}
\frac{N(E)}{D(E)}=\frac{\sum_{n=1}^m w_n F_n / (E - E_n)}{\sum_{n=1}^m w_n / (E - E_n)},
\label{aaa}
\end{equation}
where $m$ is the number of iterations of the AAA algorithm, $E_n$ are support points selected from the provided set of discrete energies, $F_n$ are the corresponding values of $F$, and $w_n$ are weights.
The polynomials $N(E)$ and $D(E)$ can be obtained respectively by multiplying the numerator and the denominator of the right-hand side of Equation~\eqref{aaa} by $\prod_{n=1}^m (E - E_n)$.
We see then that $N(E)$ and $D(E)$ are polynomials of degree $m - 1$.
The interpolation in Equation~\eqref{aaa} is exact only at the support points,
\begin{equation}
F_n=\frac{N(E_n)}{D(E_n)},
\end{equation}
and it is approximate elsewhere.
The number of iterations $m$ can be fixed by requiring the accuracy of the interpolation to be within a certain tolerance.
From Equation~\eqref{aaa}, we see that the zeros, poles, and residues of the interpolating function can be computed in terms of the support points $E_n$, the corresponding function values $F_n$, and the weights $w_n$.
A more detailed description of the AAA algorithm for rational polynomial approximation, which goes beyond the scope of this paper, can be found in Reference~\cite{AAA}.

The extrapolation of the $T$-matrix elements to complex energies via a rational polynomial approximation will produce spurious poles in addition to the physical ones.
The only poles that are deemed physical by SPARSE are those that are close enough to the input energy values that the extrapolation can be deemed stable.
A pole is discarded if its real part is outside the input energy region or too close to the either extreme of it (by default, less than $5\%$ of the input energy interval), or if the absolute value of its imaginary part is too large (by default, more than $20\%$ of the input energy interval).%
\footnote{If one is using SPARSE to calculate nominal scattering poles of the $K$-matrix, it is advisable to set the cutoff for the imaginary part as close to zero as possible. The physical poles of the $K$-matrix can only be located on the real axis.}

\section{Showcase}
\label{sec:showcase}

\begin{table}
\centering
\caption{\label{tab:channels}Integer label $i$, orbital angular momentum $l$, reduced mass $\mu$, and threshold $V^\infty$ for the channels in the SPARSE showcase example.}
\begin{tabular}{rrrr}
\toprule
$i$		& $l$	& $\mu$ [MeV]	& $V^\infty$ [MeV] \\
\midrule
1		& 1		& 1000	& 0	\\
2		& 0		& 1000	& 100	\\
\bottomrule
\end{tabular}
\end{table}

Next, we examine an application of the SPARSE algorithm to a relatively simple system with two coupled channels.
We use energy-based natural units where all physical quantities are measured in MeV.
The two channels are described in Table~\ref{tab:channels}. The potential matrix is
\begin{equation}
V(r) =
\begin{pmatrix}
V_{11}(r)	& V_{12}(r)	\\
V_{21}(r)	& V_{22}(r)	\\
\end{pmatrix}.
\end{equation}
The first diagonal element $V_{11}(r)$ is a potential well with a short-distance repulsion,
\begin{equation}
V_{11}(r) = \biggl(\frac{\kappa}{r} - E_0\biggr) \exp\bigl(-r^2/r_0^2\bigr),
\end{equation}
where $\kappa = 0.05$, $E_0 = 100$ MeV, and $r_0 = 10^{-2}$ MeV$^{-1}$.
The second diagonal element $V_{22}(r)$ is the same potential well but shifted upwards by $E_0$,
\begin{equation}
V_{22}(r) = V_{11}(r) + E_0.
\end{equation}
The offdiagonal element $V_{12}(r)=V_{21}(r)$ is Gaussian coupling potential,
\begin{equation}
V_{12}(r)= G r^2/r_0^2 \exp\bigl(-r^2 / r_0^2\bigr),
\end{equation}
where $G = 50$ MeV.
The coordinate space of $r$ is a grid of points in the interval $[0,1]$ MeV$^{-1}$ with a discretization step $d=10^{-6}$ MeV$^{-1}$.
The potential $V_{11}(r)$ has been engineered so that it has a $P$-wave (that is, $l=1$) bound state with energy somewhat below the first threshold, $V^\infty_1=0$.
The potential $V_{22}(r)$ has been engineered so that in isolation it would have an $S$-wave (that is, $l=0$) bound state with energy somewhat below the second threshold, $V^\infty_2=E_0$.
The coupling potential $V_{12}(r)$ has been engineered so that the would-be bound state in the second channel appears instead as a narrow scattering resonance in the first channel.
These potential matrix elements are plotted in Figure~\ref{fig:potentials}.

\begin{figure}
\centering
\tikzsetnextfilename{potentials}
\begin{tikzpicture}
	\begin{axis}[%
		width=7.4cm,
		xlabel={$r$},
		ylabel={$V$},
		xmin=0,
		xmax=0.05,
		ymax=220,
		x SI prefix=mega,
		x unit=eV^{-1},
		y SI prefix=mega,
		y unit=eV,
		]
		\begin{scope}
		[samples=200,
		domain=0.0001:0.05
		]
			\addplot+
			{(0.05 / x - 100) * exp(-(100 * x)^2)};
			\addplot+
			{(0.05 / x - 100) * exp(-(100 * x)^2) + 100};
			\addplot+
			{50 * (100 * x)^2 * exp(-(100 * x)^2)};
		\end{scope}
		\legend{$V_{11}$,$V_{22}$,$V_{12}$}
	\end{axis}
\end{tikzpicture}
\caption{\label{fig:potentials}Potential matrix elements for the SPARSE showcase example.}
\end{figure}

We use SPARSE to calculate the $K$-matrix for a grid of energies in the interval $[0,80]$ MeV with a constant spacing of 0.05 MeV. The $K$-matrix has only one element, since there is only one open channel in this energy region. We plot the $K$-matrix as a function of the energy in the left panel of Figure~\ref{fig:ktmatrix}. We also calculate the square modulus of the $T$-matrix element and plot it in the right panel of Figure~\ref{fig:ktmatrix}.

\begin{figure}
\centering
\tikzsetnextfilename{matrices}
\begin{tikzpicture}
	\begin{axis}[%
		at={(0,0)},
		width=6cm,
		xlabel={$E$},
		ylabel={$K$},
		xmin = 0,
		xmax = 80,
		x SI prefix=mega,
		x unit=eV,
		]
		\addplot+ table[header=false, x index=0, y index=1, col sep=comma, filter discard warning=false] {kmat.csv};
	\end{axis}
	\begin{axis}[%
	at={(11cm,0)},
	anchor=south east,
	width=6cm,
	xlabel={$E$},
	ylabel={$\lvert T\rvert^2$},
	xmin = 0,
	xmax = 80,
	x SI prefix=mega,
	x unit=eV,
	]
	\addplot+ table[header=false, x index=0, y index=1, col sep=comma, filter discard warning=false] {amp.csv};
\end{axis}
\end{tikzpicture}
\caption{\label{fig:ktmatrix}$K$-matrix (left) and square modulus of the $T$-matrix (right) in the SPARSE showcase example.}
\end{figure}

The scattering amplitude looks complicated because it results from the overlap of two different line shapes.
The smooth enhancement is due to the bound state in channel 1, which has an energy of -12.5 MeV. The sharp oscillation of the amplitude in between 60 MeV and 80 MeV is due to a resonance with mass $M=67.1$~MeV and width $\Gamma = 0.4$~MeV.

The SPARSE repository contains a Python script that prepares the CSV files \verb!channels.csv! and \verb!potential.csv! used for this example. It also contains a handy IPython notebook which reproduces the results of this section using the SPARSE Application Programming Interface.
The reader is invited to execute the IPython notebook as a tutorial for using SPARSE.

\section{Summary}
\label{sec:summary}

SPARSE is a streamlined Python algorithm for the calculation of amplitudes and resonances for elastic or inelastic scattering of particles with spin in nonrelativistic quantum mechanics.
SPARSE solves a system of coupled radial Schr\"odinger equations that represents a partial wave of a two-body scattering process in systems where the particles can change their state during the collision.
The SPARSE algorithm is optimized for the application to systems with tens of coupled Schr\"odinger equations.
It reduces the system of differential equations to a nonhomogeneous linear system by approximating the second derivative using the finite-difference method and imposing Dirichlet boundary conditions.
It calculates the numerical wavefunctions by solving the system, then compares them to the analytical wavefunctions of scattering states to obtain the $K$-matrix elements.
Scattering amplitudes for real energies are obtained from the $K$-matrix, and poles are calculated by numerical extrapolation of the scattering amplitudes into the complex plane.

The only inputs required by SPARSE are the orbital angular momenta, reduced masses, and thresholds for the each of the Schr\"odinger equations in the system plus the numerical coordinate space and potential matrix.
These numerical inputs are provided by the user in the form of two CSV files named \verb!channels.csv! and \verb!potential.csv!.
The user can prepare these numerical inputs using any computational software.
The SPARSE Application Programming Interface is designed so that it can be used with minimal Python literacy.
The SPARSE repository also contains a Python script that creates working examples for the input CSV files.
It also contains an IPython notebook which serves as tutorial for SPARSE while showcasing its functionalities.

\section*{Acknowledgements}
\noindent
This work was supported in part by the U.S. Department of Energy under grant DE-SC0011726.
I would like to thank F. Alasiri for beta testing the program.
I would also like to thank A. Rodas for discussions during the HADRON 2025 conference.

\printbibliography

\end{document}